\documentclass[aps,prb,twocolumn,showpacs,preprintnumbers,amsmath,amssymb,superscriptaddress]{revtex4}
\usepackage{dcolumn} \usepackage{bm} \usepackage{graphicx}
\usepackage{graphics} \usepackage{epsfig} \usepackage{float}
\usepackage{amsmath} \usepackage{amssymb} \usepackage{bm}
\newcommand{\degree}{\ensuremath{^\circ}}
\begin{document}

\title{Electrical Resistivity and Specific Heat of EuFe$_{2}$As$_{2}$
  Single Crystals: Magnetic homologue of SrFe$_{2}$As$_{2}$}

\author{H. S. Jeevan}
\email{jeevan@cpfs.mpg.de} \affiliation{I. Physik. Institut,
Georg-August-Universit\"{a}t G\"{o}ttingen, D-37077 G\"{o}ttingen,
Germany}

\author{Z. Hossain}
\affiliation{Department of Physics, Indian Institute of Technology,
Kanpur 208016, India}

\author{Deepa Kasinathan}
\affiliation{Max-Planck Institute for Chemical Physics of Solids,
D-01187 Dresden, Germany}

\author{H. Rosner}
\affiliation{Max-Planck Institute for Chemical Physics of Solids,
D-01187 Dresden, Germany}

\author{C. Geibel}
\affiliation{Max-Planck Institute for Chemical Physics of Solids,
D-01187 Dresden, Germany}

\author{P. Gegenwart}
\affiliation{I. Physik. Institut, Georg-August-Universit\"{a}t
G\"{o}ttingen, D-37077 G\"{o}ttingen, Germany}
%\homepage{}

\date{\today}

\begin{abstract}
We have grown single crystals of EuFe$_{2}$As$_{2}$ and investigated
its electrical transport and thermodynamic properties.  Electrical
resistivity and specific heat measurements clearly establish the
intrinsic nature of magnetic phase transitions at 20 K and 195
K. While the high temperature phase transition is associated with the
itinerant moment of Fe, the low temperature phase transition is due to
magnetic order of localized Eu-moments.  Band structure calculations
point out a very close similarity of the electronic structure with
SrFe$_{2}$As$_{2}$.  Magnetically, the Eu and Fe$_{2}$As$_{2}$
sublattice are nearly de-coupled.

\end{abstract}


\maketitle

\section{\textbf{Introduction}}

Investigations of the FeAs based compounds are presently on the rise
within the condensed matter community due to relatively high
superconducting transition temperatures of electron or hole doped
RFeAsO (R = La, Sm) and hole doped AFe$_{2}$As$_{2}$ compounds (A =
Sr, Ba) \cite{Kamihara,Takahashi,Chen,G.F.Chen, Krellner, Rotter,johrendt}. The
FeAs layers common to both series of compounds seem to be responsible for
superconductivity. Since superconductivity in this new class of
compounds coincides with the disappearance of a spin density wave (SDW) type
magnetic transition, spin fluctuations of Fe moments are suggested to
play a crucial role in establishing the superconducting ground state
in these novel superconductors \cite{Haule}.  The importance of spin
fluctuations for unconventional superconductivity has already been
realized while discussing the properties of heavy fermion
superconductors like CeCu$_{2}$Si$_{2}$, CeMIn$_{5}$ (M = Co, Rh, Ir)
as well as high T$_{c}$ cuprates \cite{Steglich}. Even though we
already have numerous superconductors encompassing the whole range of
temperature up to 160~K, the discovery of FeAs based superconductors
was greeted with great attention primarily due to its apparent
non-phononic origin and non-copper based origin.  This is the first
time that magnetic fluctuations of the "truly magnetic" element iron may
produce superconductivity at such a high temperatures.

In RFeAsO compounds, the superconducting transition temperature
increases as one substitutes La by smaller rare-earth atoms: $T_c$
rises from 28~K in LaFeAs(O,F) to 52~K in SmFeAs(O,F) \cite{Kamihara,
  Chen}. Very recently, superconductivity has also been reported in
RFe$_{2}$As$_{2}$ compounds for (K,Ba)Fe$_{2}$As$_{2}$ and
(K,Sr)Fe$_{2}$As$_{2}$ with superconducting transition temperatures as
high as 38 K. As in the case of RFeAsO, the parent compound
RFe$_{2}$As$_{2}$ also exhibits a spin-density-type phase
transition in the temperature range between 160 and 200~K for Ba and
Sr compounds which is suppressed by substituting
K for the cationic site. The
suppression of the SDW transition is accompanied by the appearance of
superconductivity \cite{G.F.Chen, Rotter}. Just like in
BaFe$_{2}$As$_{2}$ and SrFe$_{2}$As$_{2}$ the presence of a magnetic
transition at 190~K in EuFe$_{2}$As$_{2}$ [9] was reported more than a
decade ago using M\"{o}{\ss}bauer spectroscopy. From the small value
of the hyperfine field the Fe magnetism was suggested to be of
itinerant character.  Magnetic susceptibility as well as
M\"{o}{\ss}bauer spectroscopy further revealed the magnetic order of
localized Eu-moments at 20 K.  However, there are no reports of
electrical transport or thermodynamic properties on this system
yet. Because of the topical interest we have therefore decided to
prepare high-quality single crystalline samples and to investigate the
phase transitions in this compound using electrical resistivity and
specific heat measurements.

\section{\textbf{Methods}}

\subsection{\textbf{Experimental}}

Single crystals were obtained using a Bridgman method. Staring
elements (Eu 99.99$\%$, Fe 99.99 $\%$, As 99.99999$\%$) were taken in
an Al$_{2}$O$_{3}$ crucible which was then sealed in a Ta-crucible
under Argon atmosphere. The sample handling was done in a glove box
under very pure environment (oxygen $<$ 1 ppm, H$_{2}$O $<$ 1
ppm). The sealed crucible was heated under Argon atmosphere at a rate
of 30$^{\degree}$C/hour to 600$^{\degree}$C. The sample was kept at
600$^{\degree}$C for 12 hours and then heated to 900$^{\degree}$C
where again the sample was kept for 1 hour. After this the sample was
taken to 1300$^{\degree}$C, kept there for 3 hours and then slowly
cooled. We obtained several plate-like single crystals using this
process. The quality of the single crystals was checked using the Laue
method and powder x-ray diffraction as well as scanning electron
microscopy equipped with energy dispersive x-ray analysis. Electrical
resistivity and specific heat were measured using Physical Properties
Measurement System (PPMS, Quantum Design, USA). Magnetic properties
were measured using Superconducting Quantum Interference Device
(SQUID) magnetometer procured from Quantum Design, USA.

\subsection{\textbf{Theoretical}}

We have performed density functional band structure calculations using
two full potential codes: WIEN2K\cite{wien} and FPLO\cite{fplo} using
the local (spin) density approximation (L(S)DA) including spin-orbit
coupling. Additionally, we have included the strong Coulomb repulsion
in the Eu 4$f$ orbitals on a mean field level using the LSDA+$U$
approximation applying the atomic-limit double counting scheme.  We
used the Perdew-Wang\cite{perdew} flavor of the exchange correlation
potential and the energies were converged on a dense $k$- mesh with
24$^{3}$ points. We used the current experimental lattice parameters.
The As position ($z$=0.362) was taken according to
Ref. \onlinecite{Raffius}. There exists no spectroscopy data for
EuFe$_{2}$As$_{2}$, therefore we have used a $U$ of 8 eV, standard
value for a Eu$^{2+}$ ion. The results were checked for consistency
with varying $U$ values.

\section{\textbf{Results and Discussion}}

\subsection{\textbf{Experimental}}

Powder x-ray diffraction of the crushed single crystals confirms the
single-phase nature of the single crystals which form in the
ThCr$_2$Si$_2$ type tetragonal structure with lattice parameters $a$ =
3.907(4)\AA\ and $c$ = 12.114(3)\AA\, in good agreement with the
values reported in the literature \cite{Raffius}. Laue patterns reveal
that plate-like single crystals are oriented with their $c$-axis
perpendicular to the plates. The EDAX composition analysis shows that
the crystals have uniform compositions with the expected ratio of the
elements. There is a tiny percentage of EuAs binary phase in the
sample.\\

\begin{figure}[b]
\includegraphics[width=6cm,angle=270]{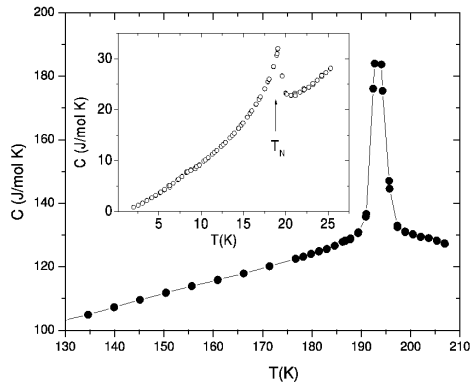}
\caption{\label{fig1} Temperature dependence of the specific heat of
  EuFe$_{2}$As$_2$ in the temperature range between 130 and 210 K.
  The inset shows the low-temperature part from 2 to 25 K. The small
  bump around 8 K is due to small amounts of EuAs binary
  phase.}
\end{figure}
\vspace{3cm}

Since the magnetic transitions of EuFe$_2$As$_2$ have already been
reported using magnetic susceptibility and M\"{o}{\ss}bauer
spectroscopy, we focus our investigation on electrical resistivity and
specific heat measurements. Fig. \ref{fig1} shows the specific heat
data taken in the temperature range between 2 and 210~K. The
low-temperature part of the specific heat shows a lambda-type anomaly
with a peak at $T_N=19$~K associated with the magnetic transition of
Eu-moments. The magnetic entropy associated with this transition is
close to $R\ln8$ as expected for $J = S = 7/2$ Eu$^{2+}$ ions. Even in
the presence of a large lattice specific heat background, a clear
signature of the high-temperature transition is observed in the form
of a very pronounced jump of the specific heat at
$T_{SDW}=195$~K. This peak reaches a maximum value of more than
50~J/mol K (on top of the phonon contribution) in spite of the small
value of the magnetic moment evidenced by the weak hyperfine field
\cite{Raffius}. The sharp peak would be compatible with a first-order
phase transition, similar as found in SrFe$_2$As$_2$
\cite{Krellner}. \\

\begin{figure}[t]
\includegraphics[width=6cm,angle=270]{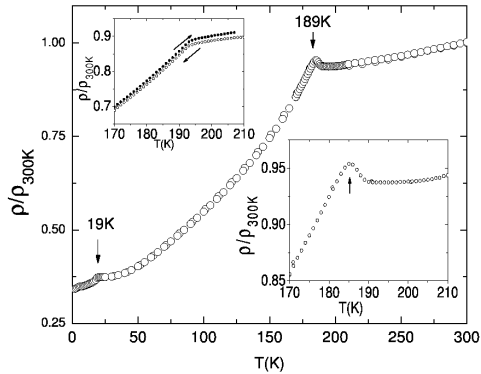}
\caption{\label{fig2} Temperature dependence of the electrical
  resistivity of EuFe$_{2}$As$_2$ measured on a plate-like single
  crystalline sample down to 2~K. The arrow indicate anomalies at 19
  and 189~K. The lower inset shows the electrical resistivity data in
  temperature range between 160 and 210~K. The upper inset shows
  corresponding data measured on a polycrystalline sample.}
\end{figure}
\vspace{3cm}

Further insight on the phase transitions is obtained from electrical
resistivity of the single crystals for the current flowing within the
$ab$-plane (Fig. \ref{fig2}). The electrical resistivity decreases marginally with
decreasing temperature down to 200~K. At $\approx$ 190~K, $\rho(T)$
shows a slight upturn leading to a peak before decreasing rapidly with
further reduction of temperature. We attribute this increase of the
resistivity to the opening of a gap at the Fermi surface due to the
formation of the SDW. Similar resistivity signatures have previously
been observed at the SDW transitions in e.g. Cr metal \cite{Rapp},
URh$_{2}$Si$_{2}$ \cite{Palstra,Uedaa} and CeCu$_{2}$Si$_{2}$
\cite{Gegenwart}. The fact that the resistivity upturn has not been
observed in polycrystalline samples implies an anisotropic temperature
dependence of the resistivity at $T_{SDW}$. Upon lowering the
temperature from $\approx 180$~K to 20 K the resistivity continues to
decrease. At 20 K a further kink is found which results from the
ordering of localized Eu-moments. The resistivity ratio
$\rho$(2K)/$\rho$(300K)$ = 3$ is fairly small but of similar magnitude
as found for many 122 compounds crystalizing in the ThCr$_{2}$Si$_{2}$
type tetragonal structure.

Since EuFe$_{2}$As$_{2}$ has similar properties compared to
BaFe$_{2}$As$_{2}$ and SrFe$_{2}$As$_{2}$, one would like to suppress
the Fe magnetism of this compound in order to induce
superconductivity. As a first approach we have tried to dope divalent
Eu with monovalent K similar to (K,M)Fe$_{2}$As$_{2}$ ( M = Ba,
Sr). Our starting composition was
K$_{0.35}$Eu$_{0.65}$Fe$_{2}$As$_{2}$. However, our results suggest
that a significant part of K evaporated out of the sample due to its
high vapor pressure and due to our high-temperature crystal growth
method, resulting is a much lower K-content in the sample. We found
clear signatures of both magnetic phase transitions in the K-doped
sample although the transition temperatures are reduced by a few
degrees and the transitions are slightly broadened (not shown)
compared to the parent compound. Obviously, we need to improve the
doping process in order to suppress the SDW and induce
superconductivity.

It is well known that the application of pressure is a 
suitable tool to change the valence state of Eu from a $f^{7}$ 
configuration to a $f^{6}$ state. This transition from Eu$^{2+}$ to 
Eu$^{3+}$ would provide an additional charge to the FeAs layer 
as an alternative way to chemical doping. 
Therefore, a study of the pressure dependence of the Eu valence is
very promising.

\subsection{\textbf{Theoretical}}

\begin{figure}[t]
\begin{center}
\includegraphics[%
  clip,
  width=8cm,
  angle=-0]{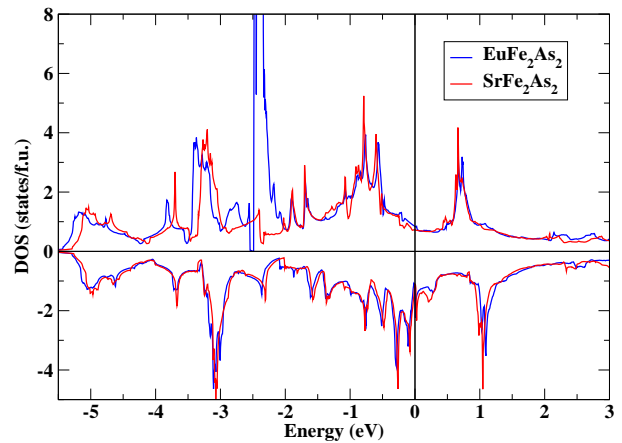}
\end{center}
\caption{\label{dos}Comparison of  the total DOS for EuFe$_{2}$As$_{2}$
  (LSDA+$U$, with ferromagnetic interaction between the Eu and Fe
  spins) and SrFe$_{2}$As$_{2}$ (LSDA). The experimental lattice was
  used for both the compounds.  The large peak for the spin-up states
  at -2.2 eV arises from the fully filled Eu 4$f$ states. The unfilled
  Eu 4$f$ states are at 9.5 eV above the Fermi level.  }
\end{figure}

To gain deeper insight into the electronic structure of
EuFe$_2$As$_2$ and to locate the system within the AFe$_2$As$_2$ family
we carried out full potential electronic structure calculations within
the L(S)DA and LSDA+$U$. Usually, for intermetallic Eu compounds, LSDA
calculations result in a Eu$^{3+}$ state with the Eu 4$f$ electrons at
the Fermi level $\varepsilon_F$. This well known flaw is in contrast
to experimental observations, caused by the underestimation of the
strong Couloumb repulsion $U$ for the localized Eu 4$f$ electrons
within the L(S)DA.  Adding the Couloumb correlation $U$ on a mean
field level using LSDA+$U$, the 4$f$ states split by about $U$ into an
occupied an an unoccupied complex. Surprisingly, already the LSDA for
EuFe$_2$As$_2$ yields a Eu$^{2+}$ state with the occupied Eu 4$f$
complex at about -0.5 eV and the unoccupied about 4 eV above
$\varepsilon_F$. This result suggests a rather stable divalent Eu
state and is confirmed by our LSDA+$U$ where the seven occupied Eu
4$f$ bands are shifted further down in energy (see Fig.~\ref{dos}).

Apart from the localized Eu 4$f$, the resulting electronic density of
states (DOS) is almost identical with the DOS of the isovalent
SrFe$_2$As$_2$ (see Fig.~\ref{dos}), especially for the Fe 3$d$ states
close to $\varepsilon_F$. Within the physically realistic range, this
result is independent of the choice of $U$. The strong similarity between
both systems holds even for details of the band structure and the related
Fermi surfaces (not shown). To study whether the small differences in
the electronic structure arise from the change of structural parameters or
from the change of the cation, we calculated the DOS for the
crystallographic data of the Eu compound where Eu was replaced by Sr.
Comparing the resulting DOS with that of SrFe$_2$As$_2$, we can assign
these differences almost exclusively to the change in crystal geometry
rather than to the substitution of the magnetic Eu$^{2+}$ ion by Sr.

In SrFe$_{2}$As$_{2}$, only the Fe atoms are magnetic, while in
EuFe$_{2}$As$_{2}$ there are 2 magnetic sub-lattices.  This raises the
question whether the ordering of the Eu moments around 20 K in
EuFe$_{2}$As$_{2}$ is aided by the interaction with the Fe spins or
just due to Eu-Eu interaction. In order to estimate the strength of
the Eu and Fe sub-lattice interaction we performed fixed spin moment
calculations within the LSDA+$U$ formalism, for 
ferromagnetic Eu and Fe sublattices.  It is energetically
favorable by 2.5 meV/f.u. to have the Eu and Fe moments parallel to
each other rather than antiparallel.  Since the Eu 4$f$ states are
quite localized, we can map the energy difference from our fixed spin
moment calculations onto a simple Heisenberg model to get an estimate
for an effective nearest neighbor exchange (J$_{eff}$) between the Eu
and Fe sub-lattice.  We obtain a value of about 1 K for J$_{eff}$ which
is rather small compared to the experimental ordering temperature for
the Eu moments ($\approx$ 20 K). This implies that the Eu and Fe
sub-lattices are quite decoupled in this system.

Furthermore, the inclusion of spin-orbit coupling allows an estimate
of the spin anisotropy for the compound. Comparing the total energies
for Fe 3$d$ spins laying the basal plane (along 100) and Fe spins
pointing along the $c$ axis (along 001) we find an easy plane
anisotropy of about 0.15 meV per Fe site.  Our result is consistent
with the magnetic structure proposed by recent neutron scattering
measurements on the magnetic homologue BaFe$_{2}$As$_{2}$\cite{huang}.
For the S=7/2 Eu$^{2+}$ ion, we obtain a negligible (easy axis)
anisotropy at the border of numerical accuracy in agreement with the
expectations for a spherical half filled 4$f$ shell.

To investigate  the possibility of a pressure induced valency change
of the Eu from 2+ to 3+, we performed a series of calculations for
reduced unit cell volumes of EuFe$_{2}$As$_{2}$. We maintained a
constant $c/a$ ratio and a constant As $z$ parameter.  For up to a 35\%
reduction of the experimental volume, we did not observe evidence for
a change in the valency of the Eu ion. This result suggests that, only
for very high pressures of the order of 50 GPa or above a valence
change of Eu might be expected.

\section{\textbf{Conclusion}}
We have grown single crystals of EuFe$_{2}$As$_{2}$ using a Bridgman
technique in a closed Ta crucible. Measurements of the electrical
resistivity and the specific heat clearly establish the magnetic
ordering of localised Eu-moments at $T_N=20$~K and itinerant Fe
moments at $T_{SDW}=190$~K.  Band structure calculations reveal a
close similarity of the electronic structure of EuFe$_{2}$As$_{2}$ and
SrFe$_{2}$As$_{2}$. The Eu and Fe$_{2}$As$_{2}$ sub-lattice are nearly
de-coupled with Fe spins preferably oriented within the $a-b$ plane.
Our preliminary attempt to dope K in place of Eu did not produce a
superconducting ground state. Further doping experiments 
are highly desirable to study possible superconductivty in this
compound and its interplay with the Eu magnetic moments.

The authors would like to thank A Leithe- Jasper for supplying the KAs
compound and U. Burkhardt and P. Scheppen for chemical analysis of the
samples. We acknowledge financial support by the DFG Research Unit 960
and BRNS (grant no. 2007/37/28).

\end{document}